\documentclass[twocolumn,aps,prl,superscriptaddress]{revtex4-2}

\usepackage{graphicx}
\usepackage[version=3]{mhchem}
\usepackage{braket}
\usepackage{ulem}
\usepackage{here}
\usepackage{dcolumn}
\usepackage{color}
\usepackage{wrapfig}
\usepackage{bm}
\usepackage{xcolor}
\usepackage{lipsum}
\usepackage{cancel}
\usepackage{soul}
\usepackage[utf8]{inputenc}
\usepackage{amsmath}

\begin{document}
\title{All-optical helicity-dependent switching in NiCo$_2$O$_4$ thin films}
\author{Ryunosuke Takahashi}  \email{bagdiners@gmail.com}
\affiliation{Department of Material Science, Graduate School of Science, University of Hyogo, Ako, Hyogo 678-1297, Japan}
\author{Yann Le Guen}
\affiliation{Institut Jean Lamour, CNRS UMR 7198, Université de Lorraine, F-54506 Nancy, France.}
\author{Suguru Nakata}
\affiliation{Department of Material Science, Graduate School of Science, University of Hyogo, Ako, Hyogo 678-1297, Japan}
\author{Junta Igarashi}
\affiliation{Institut Jean Lamour, CNRS UMR 7198, Université de Lorraine, F-54506 Nancy, France.}
\author{Julius Hohlfeld}
\affiliation{Institut Jean Lamour, CNRS UMR 7198, Université de Lorraine, F-54506 Nancy, France.}
\author{Gr\'{e}gory Malinowski}
\affiliation{Institut Jean Lamour, CNRS UMR 7198, Université de Lorraine, F-54506 Nancy, France.}
\author{Lingling Xie}
\affiliation{Institute for Chemical Research, Kyoto University, Uji, Kyoto 611-0011, Japan}
\author{Daisuke Kan}
\affiliation{Institute for Chemical Research, Kyoto University, Uji, Kyoto 611-0011, Japan}
\author{Yuichi Shimakawa}
\affiliation{Institute for Chemical Research, Kyoto University, Uji, Kyoto 611-0011, Japan}
\author{St\'{e}phane Mangin}
\affiliation{Institut Jean Lamour, CNRS UMR 7198, Université de Lorraine, F-54506 Nancy, France.}
\author{Hiroki Wadati}
\affiliation{Department of Material Science, Graduate School of Science, University of Hyogo, Ako, Hyogo 678-1297, Japan}
\affiliation{Institute of Laser Engineering, Osaka University, Suita, Osaka 565-0871, Japan}

\begin{abstract}
All-optical switching (AOS) involves manipulating magnetization using only a pulsed laser, presenting a promising approach for next-generation magnetic recording devices. NiCo$_2$O$_4$ (NCO) thin films, a rare-earth-free ferrimagnetic oxide, exhibit a high Curie temperature and strong perpendicular magnetic anisotropy. This study demonstrates AOS in NCO thin films at room temperature using long-duration laser pulses and high repetition rates. Unlike previous findings, the AOS phenomena we report here are helicity-dependent and observable with an ultrashort pulsed laser. Consequently, two distinct types of AOS can be observed in a single NCO thin film, contingent on the characteristics of the laser pulses and temperature.

\end{abstract}

\maketitle
Since the discovery of ultrafast demagnetization in Ni in 1996 \cite{Beaurepaire1996-ph}, with demagnetization times of less than 1 ps, the underlying mechanism has been extensively studied towards the high potential for next-generation non-volatile spintronics applications \cite{Kirilyuk2010-ue,El_Hadri2017-zy}. 
All-optical switching (AOS) is a phenomenon where magnetization is reversed only by laser irradiation without applying a magnetic field. This has been discovered in ferrimagnetic GdFeCo alloys \cite{Stanciu2007-kd}. Different types of AOS have been identified \cite{ElHadriPRB2016}. One is all-optical helicity-independent switching (AO-HIS) \cite{Stanciu2007-kd,Ostler2012-oz,Banerjee2020-xh,Aviles-Felix2020-uk,Radu2011-fp}, another is all-optical helicity-dependent switching (AO-HDS) \cite{ElHadriPRB2016,Lambert2014,Quessab2018,Medapalli,Choi,Parlak,Yamada2022-xw,Huang2023-np}, third is attributed to magnetization precession during anisotropy reorientation \cite{Aviles-Felix2020-uk,Peng2023-ii,Peng2024-gt}.

AO-HIS, which does not depend on the light helicity and requires only a single laser pulse, has been demonstrated for Gd-transition metal (TM) alloys, multilayers, and a Heusler alloy, namely, Mn$_2$Ru$_x$Ga \cite{Ostler2012-oz,Banerjee2020-xh}.
AO-HDS has been found in various ferrimagnets and ferromagnetic heterostructures, where the light helicity determines the orientation of the magnetization, but multiple pulses are required \cite{Lambert2014}. 

The relative importance of the inverse Faraday effect (IFE), magnetic circular dichroism (MCD), and thermal gradient-induced domain wall motion (DWM) and domain nucleation is still under heavy discussion to explain AO-HDS \cite{Quessab2018,Medapalli,Parlak,Yamada2022-xw,Huang2023-np,Jon_simulation}. In the case of materials with perpendicular magnetic anisotropy (PMA), to ensure that AOS is not obscured by multi-domain formation due to the presence of demagnetization fields, the magnetic domain size must be larger than the laser spot size. This criterion is common for all materials and all types of AOS \cite{El_Hadri2017-zy, El_Hadri2016-mv}. 

Most AOS observations have been performed on metallic materials, making the exploration of AOS in rare-earth-free materials with light magnetic constituents a promising option for lightweight and sustainable solutions.
 In recent years, NiCo$_{\bm{2}}$O$_{\bm{4}}$ (NCO) thin films have garnered attention as heavy element-free oxide materials exhibiting the
PMA \cite{Xu2022-oi,Kan2020-qo,Kan2020prb,Dho2022-bd}. 
PMA is noteworthy because of its potential for higher-density magnetic recording.
NCO thin film is a ferrimagnetic oxide with a Curie temperature ($T_{C}$) higher than 400 K when its thickness is 30 nm \cite{Shen2020,Shen2020-ml}. In NCO, Co occupies the $T_{d}$ and $O_{h}$ sites, while Ni occupies the $O_{h}$ site. The valence states were estimated to be Co$^{2+}$ ($O_{h}$ site), Co$^{3+}$ ($T_{d}$ site), and Ni$^{2+\delta}$ ($O_{h}$ site) using x-ray absorption spectroscopy and x-ray MCD measurements \cite{Kan2020prb}. The spins of the $T_{d}$-site Co and $O_{h}$-site Ni are ferrimagnetically coupled \cite{Bitla2015,Kan2020prb}, and their saturated magnetizations have been determined to be approximately 2 $\mu_{B}$ \cite{Kan2020-qo}. It was previously reported that NCO thin films realize the ultrafast demagnetization within $\sim$ 0.4 ps at 300 K, indicating that NCO films have a spin polarization as large as approximately 0.7, in line with magnetoresistance measurement \cite{Takahashi2021,Shen2020-ml}. The AOS was also observed in the NCO films. This was, however, observed only at 380 K and above.
\cite{Takahashi2023}. The absence of the AOS at room temperature, rendering it impractical, has thus hindered practical applications. 

The present work shows the effect of laser parameters such as repetition rate and pulse duration on the AOS properties of NCO thin films at room temperature. 
The AOS exhibits similarities to AO-HDS observed in Co/Pt multilayers \cite{Parlak}.
The AOS observed in the present study, which depends entirely on the helicity of pulses, differs from the AOS reported in Ref. \cite{Takahashi2023}, observed with linear polarization where the helicity dependence was only 10\%. We thus argue that two types of AOS are present in a single NCO thin film.

Epitaxial NCO thin films, with thicknesses of 26 nm and 10 nm, were grown on MgAl$_{\bm{2}}$O$_{\bm{4}}$ (001) substrates using pulsed laser deposition. During the film deposition, the substrate temperature and oxygen pressure were maintained at 315$^\circ$C and 100 mTorr, respectively. 
$T_c$ decreases with the reduction in film thickness, reaching approximately 370 K at a thickness of 10 nm. The detailed methods of the NCO sample growth can be found in Refs. \cite{Kan2020-qo,Shen2020}. 
Before the AOS experiments, we characterized the magnetic properties of the NCO thin films. The Magneto-optical Kerr effect (MOKE) signal was measured at room temperature using a He-Ne laser ($\lambda = 632.8$ nm) as the probe. 
 Figure 1 (a) displays the magnetic field evolution of the Kerr angle, clearly exhibiting the MOKE hysteresis loop of NCO thin films, indicating a coercive field of 6.5 mT and 3.9 mT for the 26-nm and 10-nm NCO thin films, respectively.
 
 \begin{figure}
\centering
\includegraphics[width=\linewidth]{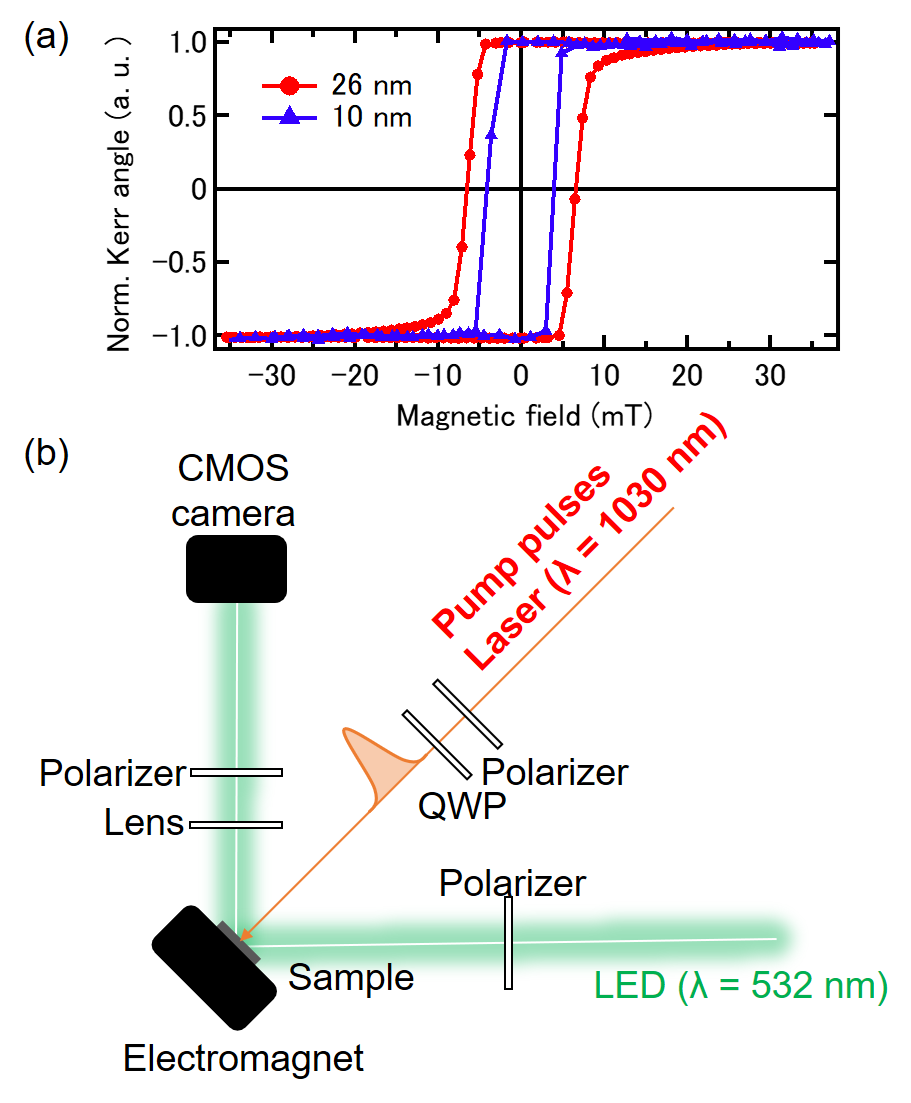}
\caption{(a) The MOKE signal as a function of the magnetic field applied perpendicular to the film plane at room temperature for 26-nm and 10-nm NCO thin films. (b) Schematics of setup to image magnetization configuration after laser pulse irradiation on the NCO thin film.}
\label{Fig1}
\end{figure}

Figure 1 (b) shows the schematics of the setup used to investigate the laser-induced effect on magnetic NCO thin films.
 A MOKE microscope is combined with a femtosecond laser system. The femtosecond laser pump pulses are focused onto the sample through a 1000 mm focusing lens, resulting in a spot size of approximately 168 \textmu m on the sample. The incident beam diameter is defined by extracting the intensity distribution from a camera-captured laser beam image, fitting a Gaussian function, and identifying the diameter at the intensity minimum of $1/e$. The linearly polarized laser pulses pass through a quarter-wave plate, resulting in right circular polarization ($\sigma^{+}$) or left circular polarization ($\sigma^{-}$) depending on the rotation of the quarter-wave plate. The circularly polarized pump pulses, with a wavelength of 1030 nm, are directed perpendicular to the film's surface. Meanwhile, a 532-nm light-emitting diode (LED) shines on the film at a 45-degree angle to the film normal to probe the magnetic configuration using a complementary metal oxide semiconductor (CMOS) camera positioned at a 90-degree angle to the probe beam. 
 The repetition rate was changed using pulse pickers equipped with laser systems that can remove pulses following the repetition rate.  Before laser pulse irradiation, NCO thin films were saturated using an external magnetic field. The laser pulse duration is determined by measuring the autocorrelation with a GECO device (Light Conversion). The full width at half maximum (FWHM) of the autocorrelation results is divided by $\sqrt{2}$ to approximate the original pulse duration \cite{Hache1996-kd}. 

First, we investigated the dependence of laser repetition rate. After saturating the NCO thin films, we irradiated the sample with $10^5$ laser pulses under a zero-applied magnetic field using a 1- and 20 kHz repetition rate and recorded the MOKE images. Figure 2 shows the MOKE images taken after irradiating the 26-nm NCO thin film with $10^5$ pulses, a $\sigma^{+}$ circular helicity, for 0.22- ps and 12.9-ps pulse duration, and 1 kHz and 20 kHz. \textit{M$^+$} indicates when magnetization is saturated up and \textit{M$^-$} when it is saturated down. The MOKE images are taken both before and after the laser irradiation.
 We subtracted the background using the pre-laser images as a reference to enhance the clarity of the post-laser images. This process involved subtracting the pixel values of the pre-laser images from the corresponding pixel values of the post-laser images, then normalizing the images by dividing the background-subtracted images by the difference between the saturated MOKE images in the \textit{M$^+$} and \textit{M$^-$} states.

\begin{figure}[H]
\centering
\includegraphics[width =\linewidth]{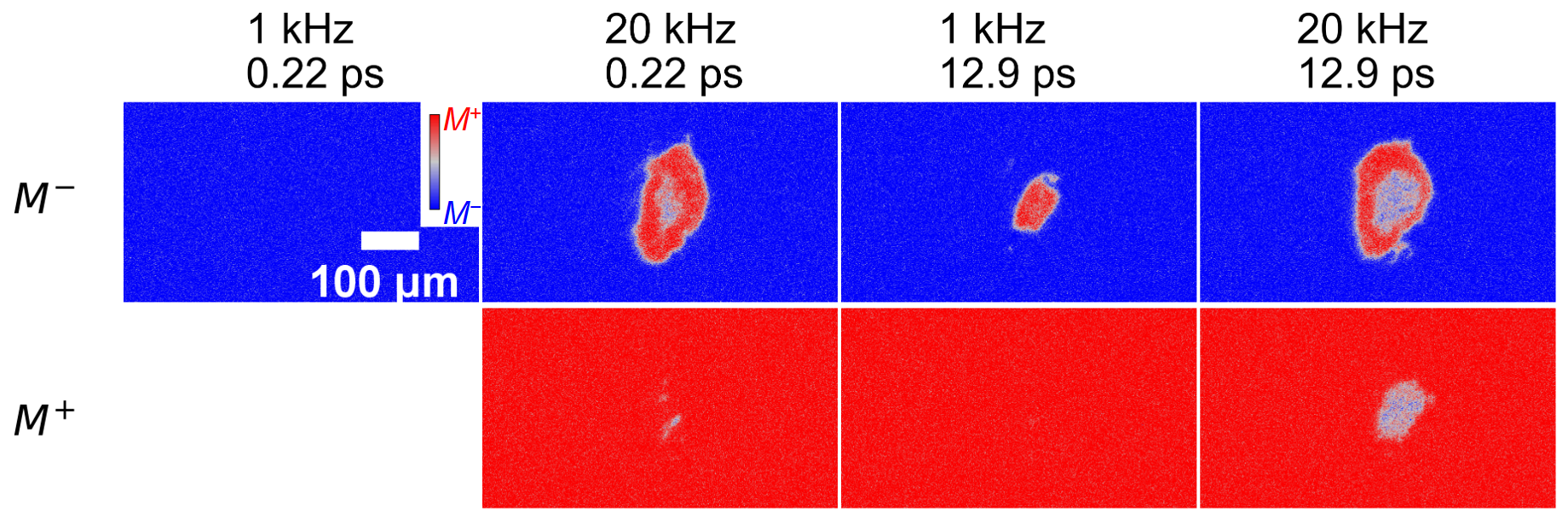}
\caption{
The MOKE images in NCO thin films after laser accumulation of $10^5$ pulses with 1 and 20 kHz repetition rates, pulse duration of 0.22 and 12.9 ps in 26-nm NCO thin films. The fluency of pump pulses was 4.0 mJ/cm$^2$.}
\label{Fig2}
\end{figure}

\begin{figure*}[htbp]
\centering
\includegraphics[width =\linewidth]{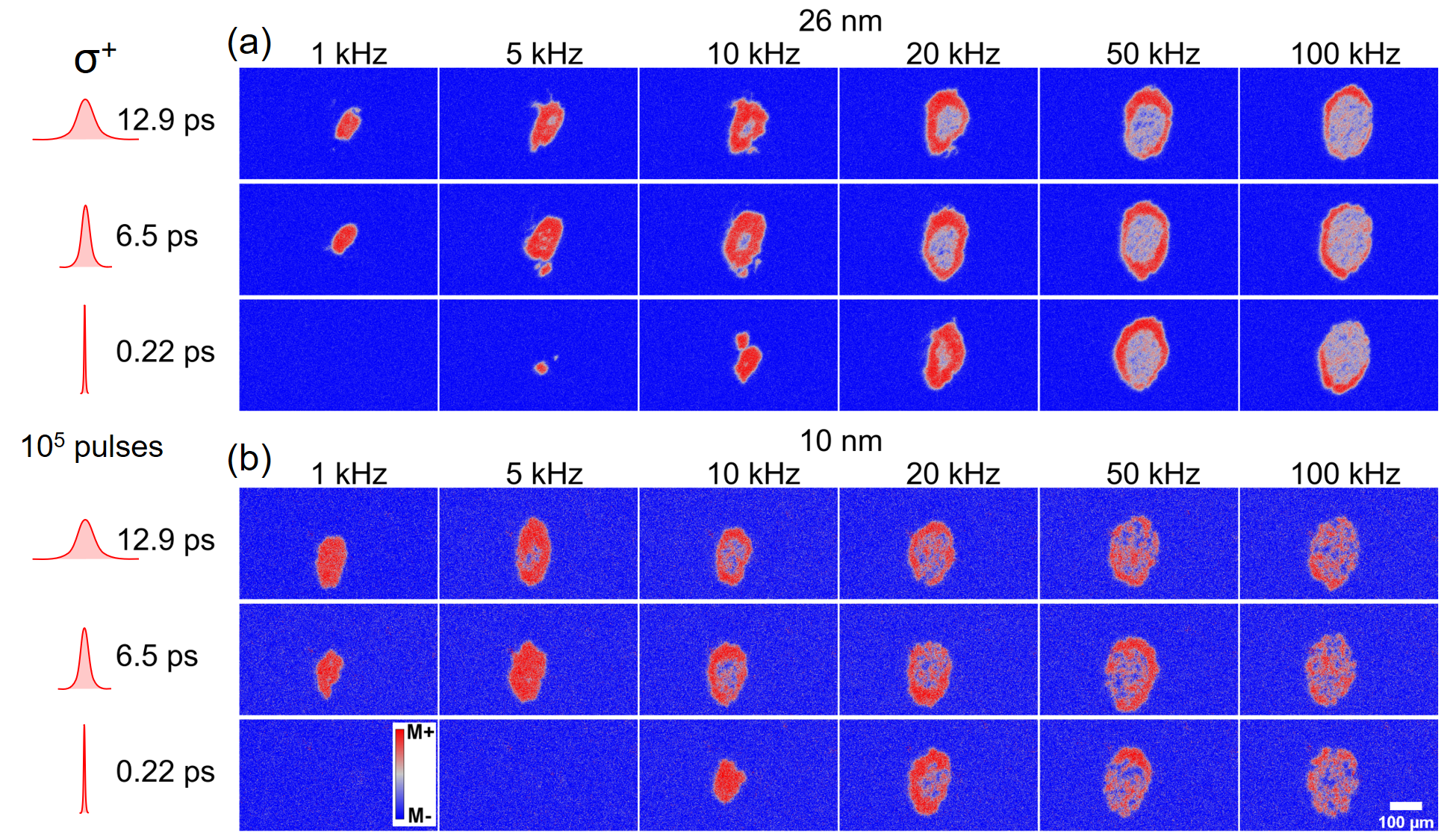}
\caption{The MOKE images after laser-pulse accumulation with various pulse durations of 0.22 ps, 6.5 ps, and 12.9 ps and repetition rate from 1 to 100 kHz in (a) 26-nm and (b) 10-nm NCO thin films using fluence of pump pulses of 4.0 mJ/cm$^2$ and 5.2 mJ/cm$^2$, respectively.}
\label{Fig3}
\end{figure*}

As shown in Fig. 2, AOS is observed shinning $10^5$ pulses with $\sigma^+$ circularly polarization and laser fluence of 4 mJ/cm$^2$ only if we are starting from the {\sl M$^{-}$} state. When the laser is irradiated in the {\sl M$^{+}$} state, AOS is never observed; only a multidomain state is created in the center of the area irradiated by the laser. From those measurements, we can conclude that AO-HDS is at room temperature in NCO thin films. Moreover, controlling laser parameters allows for observing AOS without multidomain states, as exemplified by the 1 kHz, 12.9 ps laser conditions shown in Fig. 2. Please refer to the supplementary material for the results obtained by irradiating the 10-nm and 26-nm NCO thin films with $\sigma^{-}$ polarized light at $M^{-}$ and $M^{+}$.

To comprehensively investigate the pulse width and repetition rate dependence in NCO thin films, we conducted experiments using three pulse widths—0.22 ps, 6.5 ps, and 12.9 ps—and applied pulses with repetition rates ranging from 1 kHz to 100 kHz to 26-nm and 10-nm NCO thin films. All experiments were initiated from the {\sl M$^-$} state and conducted at room temperature. Each sample was subjected to 10$^5$ pulses of right circularly polarized light $\sigma^{+}$.
 Figure 3 shows that the threshold for AOS generation varies with the pulse duration. Additionally, three distinct states were observed depending on the repetition rate: (1) AOS only, (2) AOS forming a ring with multidomain states emerging in the central region, and (3) a state where AOS no longer forms a ring, with the majority of the region transitioning to a multidomain state. For example, in 26-nm NCO thin films with a pulse duration of 12.9 ps, only AOS was observed at 1 kHz. At repetition rates between 5 and 20 kHz, an AOS ring formed with multidomains emerging in the central region. At 50 kHz and 100 kHz, AOS no longer formed a ring, and the size of the multidomains increased, with nearly the entire region becoming multidomain at 100 kHz.
The dependence on film thickness manifested in the morphology of the multidomains. In the 26-nm thin film, the multidomains appeared as white regions with a contrast value 0.5. Conversely, observing distinct up and down magnetic domains as blue and red regions in the 10-nm thin film rather than as white regions leads to the conclusion that the equilibrium domain size in the 10-nm film is larger. This larger domain size results from the film's more minor magnetic moment and lower coercive field at room temperature than the 26-nm film. Thus, the multidomain size exceeding the spatial resolution allows these distinct domains to be observed.
 These results lead us to conclude that while longer pulse durations and higher repetition rates facilitate the generation of AOS, exceeding certain thresholds favors the formation of multidomain states over AOS.
Additionally, longer pulse duration can cause a slower, more gradual heating process, allowing for more extensive thermal diffusion, consistent with previous studies \cite{Verges2024-oe}.
This AO-HDS at room temperature differs from the earlier observation of AOS in NCO thin films, where we have to raise the sample temperature to 380 K and above to observe AOS by irradiating pulses with a 1-kHz repetition rate \cite{Takahashi2023}. 
Moreover, let us discuss the heat-gradient-driven DWM velocity. Since AO-HDS cannot be observed at 1 kHz using 0.22-ps pulse duration, the DWM velocity is expected to be greater than that of [Co/Pt]$_3$ multilayer\cite{Parlak}. NCO thin film has a small coercive field of less than 10 mT, which causes the fast domain-wall motion \cite{Metaxas2007-rk} due to the gradient created by heat accumulation of Gaussian-laser pulses \cite{Parlak}.

Next, compare AO-HDS in NCO thin films with previous work reported in Ref.~\cite{Takahashi2023}. This time, we observed AO-HDS by accumulating $10^5$ pulses at room temperature. The helicity of laser pulses determines AO-HDS creation. Moreover, we need to irradiate the sample with laser pulses with a high repetition rate of more than 5 kHz when using a pulse duration of 0.22 ps. On the other hand, AOS in NCO thin films, as reported in 
Ref.~\cite{Takahashi2023}, was observed at temperatures of 380 K and above with a 1 kHz repetition rate and pulse duration of $\sim$ 0.2 ps by accumulating more than 1000 pulses with linear polarization.  
\begin{figure}
\centering
\includegraphics[width =\linewidth]{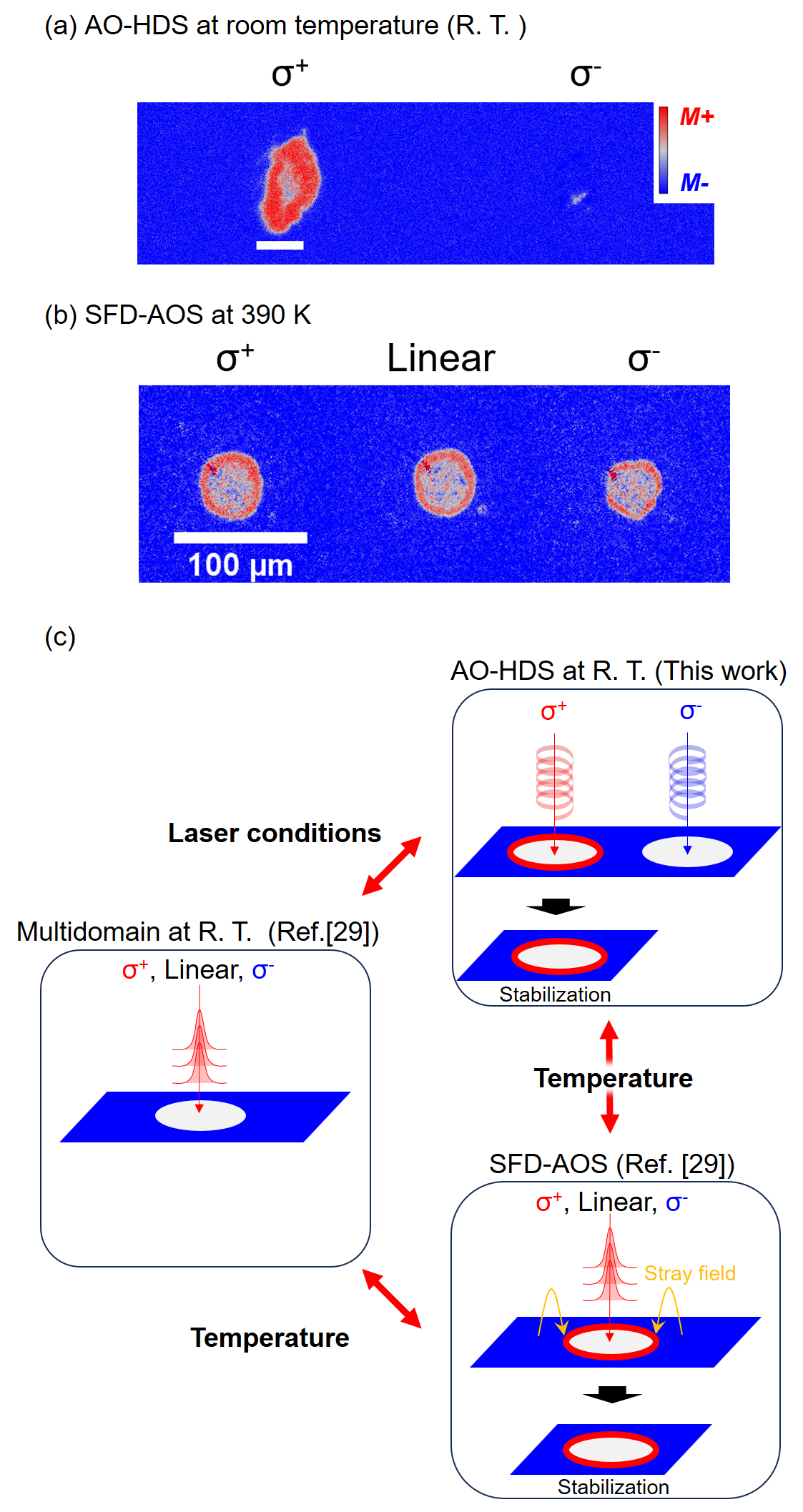}
\caption{(a) The MOKE images after 10$^5$ pulses accumulation with using the repetition rate of 20 kHz, pulse duration of 0.22 ps at room temperature in 26-nm NCO in films. (b) The MOKE images after 10$^6$-laser-pulse accumulation using the repetition rate of 30 kHz, pulse duration of $\sim$ 0.18 ps at 390 K in 26-nm NCO thin films. (c) Schematics of two types of AOS in NCO thin films.}
\label{Fig4}
\end{figure} 
To compare the AO-HDS with previously observed AOS as reported in \cite{Takahashi2023}, we observed both phenomena in 26-nm NCO thin films as shown in Fig. 4 (a, b). Figure 4 (a) shows the AO-HDS after accumulating $10^5$ pulses with $\sigma^{+}$ or $\sigma^{-}$ at room temperature only starting from {\sl M$^{-}$}. The repetition rate was 20 kHz, and the pulse duration was 0.22 ps. The complete helicity dependence was observed. Figure 4 (b) shows AOS after accumulating $10^6$ pulses with the polarization of $\sigma^{+}$, linear, or $\sigma^{-}$ at 390 K only starting from {\sl M$^{-}$}. The repetition rate and the pulse duration were 30 kHz and $\sim$ 0.18 ps, respectively.  The observed AOS contains a finite but weak helicity dependency, while the switching ring can be observed using any polarization. These results indicate minimal helicity dependence, suggesting that the growth of AOS is predominantly driven by the effects of the stray field rather than by MCD or IFE as helicity-dependent effects. The edge effects resulting from the MCD are fully helicity-dependent and are believed to contribute to the AO-HDS observed at room temperature. In addition, strong heating from the center of the laser leads to the stochastic formation of helicity-independent random domains, consistent with Ref.~\cite{Medapalli}. Although the AOS observed at 390~K exhibits weak helicity dependence due to the MCD effect, the energy barrier between up and down spins is small near \(T_c\), making the helicity dependence less pronounced compared to the room temperature AO-HDS. In this temperature regime, AOS at the perimeter of multidomain solely due to the MCD effect does not occur, necessitating the consideration of other effects, such as stray fields.

Consequently, we propose designating this phenomenon as stray field-driven all-optical switching (SFD-AOS) to distinguish it from the AO-HDS observed in this study. Figure 4 (c) shows the schematics of AOS in NCO thin films. Two types of AOS exist: AO-HDS at room temperature and SFD-AOS at high temperature. These two types of AOS are contingent upon the temperature and laser conditions, and if the conditions deviate, only the multidomain will be observed. 
In the case of SFD-AOS, helicity-dependent effects and stray fields play a role in the process, leading to a small helicity dependence of about 10\% \cite{Takahashi2023}. This finding could potentially reshape our understanding of domain growth mechanisms. According to Ref.~\cite{Gorchon2014-nr}, 
the DWM velocity increases at high temperatures under an external magnetic field \cite{Gorchon2014-nr,Igarashi2024-nx}. Therefore, the SFD-AOS observed at NCO thin films at temperatures of 380 K and above differs from the AO-HDS observed in this work, as domain shrinkage caused by DWM due to the temperature gradient is not the main factor for domain shrinkage in SFD-AOS. At the same time, a higher repetition rate is necessary to observe AO-HDS in NCO thin films. SFD-AOS can be observed even by using a low repetition rate of 0.02 kHz as reported in \cite{Takahashi2023}

Our research explored parameters such as pulse duration and repetition rate to achieve room-temperature AOS in NCO thin films. As a result, AO-HDS was observed at room temperature in NCO thin films. High repetition rates and long pulse durations were preferable for realizing AO-HDS. By comparing these results with the previous study reported in \cite{Takahashi2023}, we concluded that two types of AOS can be observed in NCO thin films. In 26-nm NCO thin films, we also observed AOS at 390 K, consistent with the previous work \cite{Takahashi2023}, where AOS was observed using any polarization. These results lead us to name this phenomenon SFD-AOS, based on the scenario that the stray field, as a helicity-independent effect, plays a dominant role in realizing the AOS ring. This study has revealed that NCO thin films, as a single material, can achieve two types of AOS depending on the laser conditions and temperature, marking the first instance of such capability.
\section*{Supplementary Material}
The supplementary material includes MOKE images of 10-nm and 26-nm NiCo$_2$O$_4$ thin films after $10^5$ pulses (0.22 ps pulse duration) of laser irradiation, showing switching from both M$^+$ and M$^-$ states, illustrating the thickness dependence of all-optical helicity-dependent switching.

\section*{Acknowledgment}
We want to thank Jon Gorchon for his valuable discussions and expert assistance with the experiment. This work was supported by JSPS KAKENHI under Grant Nos.19H05816, 19H05823, 19H05824, 21H01810, 23H01108, and 23K25805 KAKENHI and the
MEXT Quantum Leap Flagship Program (MEXT Q-LEAP) under Grant No. JPMXS0118068681. The ANR also supported this work through the SLAM project (ANR-23-CE30-0047) and the France 2030 government grants PEPR Electronic EMCOM (ANR-22-PEEL-0009) and PEPR SPIN (ANR-22-EXSP-0002), as well as the MAT-PULSE-Lorraine Université d’Excellence project (ANR-15-IDEX-04-LUE). The work was also supported by the Japan Science and Technology Agency (JST) as part of the Adopting Sustainable Partnerships for Innovative Research Ecosystem (ASPIRE) Grant JPMJAP2314. This work was partly supported by grants for the Integrated Research Consortium on Chemical Sciences and the International Collaborative Research Program of the Institute for Chemical Research in Kyoto University from Japan's Ministry of Education, Culture, Sports, Science, and Technology (MEXT).  

The corresponding author can obtain the data supporting this study's findings upon a reasonable request.

%

\end{document}